\documentclass[journal=jctcce,manuscript=article,layout=twocolumn]{achemso}
\setkeys{acs}{articletitle=true}
\SectionNumbersOn

\usepackage[utf8]{inputenc}
\usepackage[T1]{fontenc}
\usepackage{mathptmx}
\usepackage[scaled=0.92]{helvet}
\usepackage{courier}
\usepackage{mathtools}
\usepackage{amsfonts}
\usepackage{bm}
\usepackage{booktabs}
\usepackage{placeins}
\usepackage{array}
\usepackage{multirow}
\usepackage{algorithm}
\usepackage{algpseudocode}
\usepackage{siunitx}
\usepackage{xspace}
\usepackage{xcolor}
\usepackage{xurl}
\usepackage[hidelinks]{hyperref}
\DeclareMathOperator*{\argmax}{arg\,max}
\usepackage{graphicx}
\usepackage{booktabs}
\usepackage{makecell}
\usepackage{threeparttable}

\newcommand{\smin}{S$_1$-min}
\newcommand{\cioo}{\ensuremath{\mathrm{CI}_{\mathrm{OO}}}\xspace}
\newcommand{\cicc}{\ensuremath{\mathrm{CI}_{\mathrm{CC}}}\xspace}

\definecolor{color1}{RGB}{255,0,0}

\definecolor{color2}{RGB}{0,0,255}

\definecolor{color3}{RGB}{0,255,0}

\title{Toward Self-Driving Multi-reference Calculations: Belief-Aware Orbital Selection}

\author{Chen Li}
\affiliation{Computer Network Information Center, Chinese Academy of Sciences, Beijing 100190, China}
\alsoaffiliation{University of Chinese Academy of Sciences, Beijing 101408, China}
\author{Kaifan Pan}
\affiliation{Computer Network Information Center, Chinese Academy of Sciences, Beijing 100190, China}
\alsoaffiliation{University of Chinese Academy of Sciences, Beijing 101408, China}
\author{Xinyu Sun}
\affiliation{School of Chemistry and Chemical Engineering, Shandong University, Qingdao 266237, China}
\author{Yinxuan Song}
\affiliation{Ganjiang Innovation Academy, Chinese Academy of Sciences, Ganzhou 341119, China}
\author{Lianhua He}
\affiliation{Computer Network Information Center, Chinese Academy of Sciences, Beijing 100190, China}
\alsoaffiliation{National Artificial Intelligence Application Pilot Base (Science Sector, LLM Subsector), Beijing 100190, China}
\author{Teng Lu}
\affiliation{Computer Network Information Center, Chinese Academy of Sciences, Beijing 100190, China}
\alsoaffiliation{National Artificial Intelligence Application Pilot Base (Science Sector, LLM Subsector), Beijing 100190, China}
\author{Haibo Ma}
\affiliation{School of Chemistry and Chemical Engineering, Shandong University, Qingdao 266237, China}
\author{Zhenggang Lan}
\affiliation{School of Environment, South China Normal University, Guangzhou 510006, China}
\email{zhenggang.lan@m.scnu.edu.cn / zhenggang.lan@gmail.com}
\author{Yingjin Ma}
\affiliation{Computer Network Information Center, Chinese Academy of Sciences, Beijing 100190, China}
\alsoaffiliation{National Artificial Intelligence Application Pilot Base (Science Sector, LLM Subsector), Beijing 100190, China}
\email{yingjin.ma@sccas.cn}

\keywords{active-space selection, sequential decision making, partially observable Markov decision process, quadratic unconstrained binary optimization, density matrix renormalization group, multistate electronic structure}

\usepackage{etoolbox}
\usepackage{flushend}
\usepackage{tabularx}
\usepackage[protrusion=true,expansion=false]{microtype}
\makeatletter
\renewcommand{\normalsize}{%
  \@setfontsize\normalsize{10bp}{11.05bp}%
  \abovedisplayskip=6pt plus 2pt minus 2pt
  \belowdisplayskip=6pt plus 2pt minus 2pt
  \abovedisplayshortskip=3pt plus 1pt minus 1pt
  \belowdisplayshortskip=4pt plus 1pt minus 1pt}
\renewcommand{\small}{\@setfontsize\small{9bp}{10.5bp}}
\renewcommand{\footnotesize}{\@setfontsize\footnotesize{9bp}{10.5bp}}
\normalsize

\@ifpackageloaded{ctex}{}{}
\renewcommand{\section}{\@startsection{section}{1}{\z@}%
  {-11pt plus -2pt minus -1pt}{5pt}%
  {\normalfont\sffamily\bfseries\fontsize{10bp}{12bp}\selectfont}}
\renewcommand{\subsection}{\@startsection{subsection}{2}{\z@}%
  {-9pt plus -2pt minus -1pt}{4pt}%
  {\normalfont\bfseries\fontsize{10bp}{11.05bp}\selectfont}}
\renewcommand{\subsubsection}{\@startsection{subsubsection}{3}{\z@}%
  {-7pt plus -1pt minus -1pt}{-0.5em}{\normalfont\itshape}}
\let\acs@section\section
\let\acs@subsection\subsection
\let\acs@subsubsection\subsubsection
\renewenvironment{abstract}{%
  \par\noindent\textbf{\abstractname.}\enspace\ignorespaces
}{\par\vspace{6pt}}
\let\ps@acs\ps@plain
\AtBeginDocument{%
  \singlespacing\normalsize
  \setlength{\jot}{3pt}%
  \setlength{\parindent}{3.3mm}%
  \emergencystretch=1em
  \pagestyle{plain}%
  \renewcommand{\bibfont}{\fontsize{9bp}{10.5bp}\selectfont}%
  \setlength{\bibsep}{0pt plus 0.3pt}%
  \setlength{\bibhang}{1.4em}%
}
\makeatother
\begin{document}

\begin{abstract}
We reformulate active-orbital selection as a partially observable Markov decision process (POMDP), analogous to decision-making in autonomous driving and robot control, and couple it with quadratic unconstrained binary optimization (QUBO) to emulate expert-like selection of active orbitals. The three popular types of orbital information, i.e. occupation numbers, orbital composition, and orbital entanglements, can be used separately or jointly. QUBO constructs complete candidate spaces under fixed electron and orbital counts, while the POMDP selects successive calculations, updates its belief from each completed result, and thereby corrects and refines the active orbitals. This belief-aware sequential procedure enables continuous active-space selection rather than one-shot threshold-based filtering, so the space can adapt as the calculation proceeds.
We demonstrate the approach for valence-space construction in small molecules, spectral calculation of metal compound TiF$_3$, and active-space refinement of 1,2-dioxetanone around conical intersections.
Because selection learns from calculations on the molecule of interest without requiring a pre-existing training set, it is suited to following the entire course of complex chemical reactions. 
\end{abstract}

\includegraphics[scale=0.65,bb=20 0 350 180]{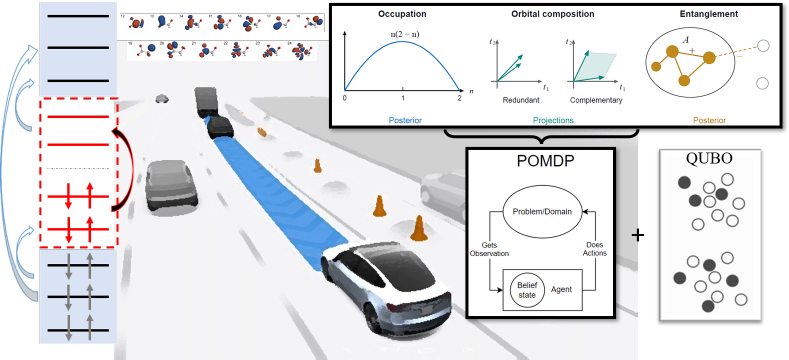}

\section{Introduction}

Multi-configurational (MC) methods describe strongly correlated electronic states in which several configurations contribute substantially, as encountered in bond dissociation, radicals, and transition-metal compounds. Active orbitals are the selected orbitals among which electrons can redistribute; together with their assigned electrons, they define the active space. Complete-active-space self-consistent field (CASSCF) includes all allowed configurations within this space and optimizes their coefficients and the orbitals.\cite{Roos1980CASSCF,Werner1985CASSCF} This captures static correlation associated with near-degenerate configurations. Multi-reference (MR) treatments, such as complete-active-space second-order perturbation theory (CASPT2)\cite{Andersson1990CASPT2} and $n$-electron valence state perturbation theory (NEVPT2),\cite{Angeli2001NEVPT2} add dynamical correlation to this reference.


The restricted-excitation methods such as restricted-active-space (RASSCF)\cite{malmqvist2002restricted, krausbeck2014choosing} and selected CI \cite{Booth2009FCIQMC, LiManni2016FCIQMCSCF,  liu2016ici, zhang2020iterative} enlarge the active space relative to CASSCF. Approaches based on reduced or compressed configurations, particularly the density-matrix renormalization group (DMRG) \cite{White1992DMRG, Chan2011DMRG, White1999QuantumChemistry, Chan2002DMRG, Legeza2003DBSS, Sharma2012SpinDMRG, Wouters2014CheMPS2, Keller2015MPO, Ren2020MPO} and neural-network quantum states (NNQS),\cite{Carleo2017NNQS, Hermann2020DeepNeural, li2024improved, ma2024quantum, kan2025nnqs} further expand the practical active-space size. Recent applications include transition-metal complexes and clusters,\cite{Marti2008TransitionMetals, Kurashige2009DMRG, Tian2022MixedPrecision} the Mn$_4$CaO$_5$ cluster,\cite{Kurashige2013MnCluster} and iron-sulfur clusters, \cite{Sharma2014FeS} which are already exceed the capacity of standard CASSCF. As accessible active spaces grow, selecting orbitals with the required chemical character and correlations becomes increasingly important.\cite{Wouters2014DMRGReview,Baiardi2020DMRGReview}

Three types of orbital information are commonly used to guide active-space selection. Occupation statistics identify departures from a closed-shell description and guide natural-orbital and generalized-valence-bond selection.\cite{Ma2013NaturalOrbitals,Zou2020GVBActiveOrbitals,Toth2020ActiveOrbitalSelection} Atomic or fragment projections identify chemical character, as in atomic-valence active space (AVAS) and imposed-CAS methods.\cite{Sayfutyarova2017AVAS,Lei2021iCAS} Single-orbital entropies measure correlation with the remaining orbitals, and mutual information measures pairwise orbital correlation.\cite{Legeza2003Entropy,Rissler2006OrbitalMI,Boguslawski2012Entanglement, Ding2023QICAS} These measures underpin {\sc autoCAS}.\cite{Stein2016AutoCAS,Stein2019AutoCASProgram} The three criteria describe complementary properties, e.g. an orbital can have the desired atomic character but weak correlation, while individually important orbitals can supply redundant information, 
so that selection must assess how orbitals contribute together within a limited orbital space.\cite{Wardzala2026MultireferenceReview}
Orbital quantum entanglement provides a direct observable of inter-orbital interactions in the view of electronic correlation, while other property-related observables can also serve as selection criteria. For example, the ABC scheme \cite{Bao2018AutomaticActiveSpace}, rank orbitals through approximate pair coefficients,\cite{King2021RankedOrbital} compare trial spaces using dipole moments \cite{Kaufold2023DipoleActiveSpace} or variational multi-configuration pair-density functional theory (MC-PDFT) energies. \cite{King2023VariationalActiveSpace} Additionally, supervised models also predict active-space membership from orbital descriptors or correlated reference data. \cite{Jeong2020MLActiveSpace, Golub2021MLTransitionMetal}

For reactions, orbital consistency across geometries is also required: bond rearrangement alters orbital character, so inconsistent choices unbalance correlation recovery and distort relative energies. \cite{Wardzala2026MultireferenceReview}
Orbital mapping addresses this problem by constructing corresponding spaces along reaction paths. \cite{Bensberg2023CorrespondingActiveSpaces}, while the weighted active-space protocol assigns consistent spaces across configurations. \cite{Seal2025WeightedActiveSpace} 
Related work on quantum embedding maintains consistent orbital partitions in systems with near-degenerate or de-localized orbitals. \cite{Kolodzeiski2023ActiveSpaceEmbedding}
These developments address continuity explicitly. A complementary task is to decide which
candidate calculation should be performed next when the active-space size and calculation budget are fixed.

That decision must account for incomplete diagnostic information and changes caused by orbital optimization. Here, we use each candidate result to update predictions for the remaining spaces and to plan the next calculation. Both accepted and rejected valid results provide information about
the target molecule. The motivation parallels the separation of sensing, state estimation, and planning in autonomous systems:\cite{Urmson2008Boss, talpes2020compute, zablocki2022explainability} observations inform a decision, and its outcome informs the next one.

We formulate this iterative refinement as a partially observable Markov decision process (POMDP),\cite{Kaelbling1998POMDP} coupled with quadratic unconstrained binary optimization (QUBO).\cite{Glover2022QUBOTutorial} 
A Bayesian posterior (i.e. ``belief'') is constructed over uncertain active spaces with indicators of orbital statistics, in which the orbital occupation, composition, and entanglement criteria can be used individually or jointly. After that, QUBO assembles candidate spaces at fixed electron and orbital counts. 
Over a horizon of at most two calculations, the POMDP selects the candidate that maximizes the expected improvement in the retained orbital space's quality. This value includes the effect of the first result on a possible second choice. Valid scores update the belief even when the candidate space is rejected.
We examine the POMDP-QUBO selection approach via the valence-space construction in small molecules, spectral calculation of metal compound TiF$_3$, and active-space refinement of 1,2-dioxetanone near conical intersections. 

The structure of the paper is organized as follows: Section~2 presents the selection framework, Section~3 gives computational settings, Section~4 discusses results, and Section~5 summarizes the conclusions.

\section{Theory and Method}\label{sec:theory}

\subsection{The active-space selection problem}\label{sec:mr_foundation}

A complete active space divides the orbitals into inactive, active, and external sets. Inactive orbitals remain doubly occupied and external orbitals remain empty. The active space $A$ contains $K$ spatial orbitals and $N_e$ electrons. Each electronic state is a linear combination of all configurations allowed by the electron count, spin, and chosen symmetry:
\begin{equation}
 |\Psi_r(A)\rangle=\sum_I c_{Ir}|\Phi_I(A)\rangle.
 \label{eq:mr_expansion}
\end{equation}
Here $r$ labels the electronic state, $I$ labels a Slater determinant or spin-adapted configuration state function with the common inactive occupation, and $c_{Ir}$ is its coefficient in the normalized state. 

\begin{figure}[htpb!]
\centering
\includegraphics[scale=0.375,bb=50 0 700 350]{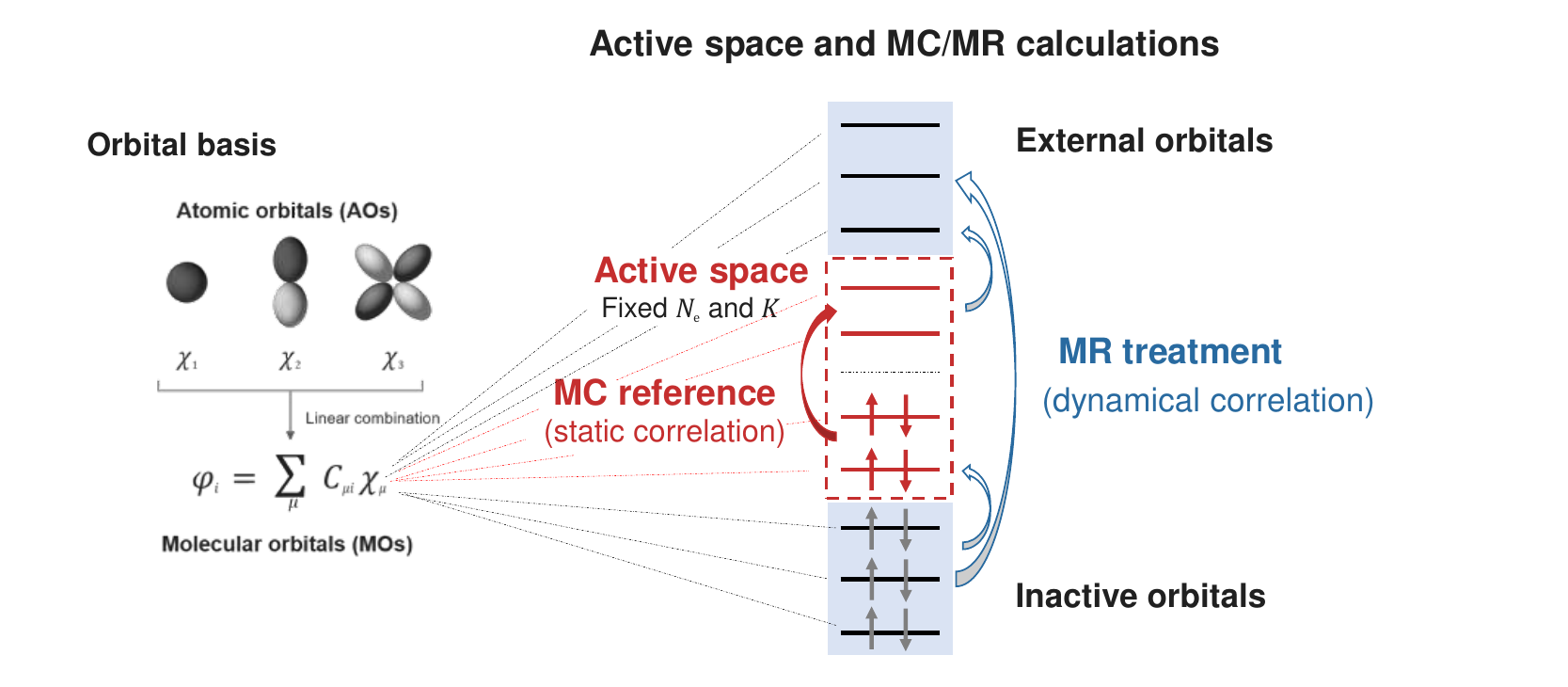}
\caption{From atomic orbitals to active-space MC/MR calculations. The MC reference describes static correlation, and subsequent MR treatment adds dynamical correlation.}
\label{fig:active_space_correlation}
\end{figure}

Truncated full configuration interaction (FCI) or CASCI optimizes the configuration coefficients in fixed orbitals. The general MCSCF approaches, which contains CASSCF,\cite{Roos1980CASSCF, Werner1985CASSCF} DMRG-SCF,\cite{Ghosh2008DMRGSCF, Zgid2008DMRGSCF, Ma2017SecondOrder} and stochastic CASSCF\cite{LiManni2016FCIQMCSCF} etc. 
also optimizes the orbitals. During the MCSCF calculation, one usually keep $N_e$ and $K$ fixed
while changing which orbitals are active. Figure~\ref{fig:active_space_correlation} illustrates
the choice of active orbitals determines the composition of the reference configurations.
Although orbital optimization is routinely performed, a suitable reference space still requires
a chemically reasonable orbital set as a prerequisite. Since the common Newton-type or super-CI type orbital-optimization methods solve the fourth-order orbital problem only approximately to at most second order, so the initial orbital choice affects both convergence and accuracy. \cite{werner1980quadratically, roos1983multiconfigurational, werner1985second, li2016combining, lischka2018multireference, kreplin2019second}

\subsection{POMDP for active space selection} \label{sec:pomdp}

\begin{figure}[htbp!]
\centering
\includegraphics[scale=0.875,bb=0 0 300 310]{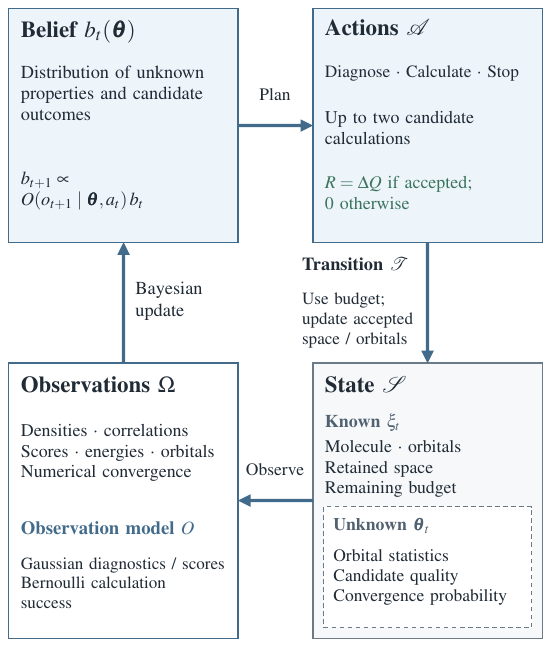}
\caption{Belief, actions, and observations in active-space selection. Symbols follow Table~\ref{tab:pomdp}.}
\label{fig:pomdp_elements}
\end{figure}

Diagnostics estimate orbital properties, and candidate calculations evaluate proposed active spaces. We describe the choice of calculation with seven POMDP elements:\cite{Kaelbling1998POMDP}
\begin{equation}
 \mathcal M=(\mathcal S,\mathcal A,\mathcal T,\Omega,O,R,\gamma),
 \label{eq:pomdp_tuple}
\end{equation}
Table~\ref{tab:pomdp} relates these elements to active-space selection. Figure~\ref{fig:pomdp_elements} shows how calculated observations update the belief and guide the next action.

\begin{table}[htbp!]
\centering
\caption{POMDP elements for active-space selection.}
\label{tab:pomdp}
\footnotesize
\renewcommand{\arraystretch}{1.04}
\setlength{\tabcolsep}{4pt}
\begin{tabularx}{\columnwidth}{@{}>{\raggedright\arraybackslash}p{0.24\columnwidth}>{\raggedright\arraybackslash}X@{}}
\toprule
Element & Meaning in active-space selection \\
\midrule
State space $\mathcal S$ & \textbf{Known:} molecule, current orbitals, retained space, and remaining budget.\newline
\textbf{Unknown:} orbital statistics, candidate quality scores, and convergence probabilities. \\
\addlinespace[3pt]
Actions $\mathcal A$ & Diagnose the orbital window, calculate a candidate active space, or stop selection. \\
\addlinespace[3pt]
Transition $\mathcal T$ & Each calculation consumes budget. Acceptance replaces the retained space and updates the orbitals when optimized. \\
\addlinespace[3pt]
Observation space $\Omega$ & Diagnostic densities and correlation measures; candidate quality scores, energies, and orbitals; convergence status. \\
\addlinespace[3pt]
Observation model $O$ & Distribution of results for a given state and action: Gaussian models for diagnostic uncertainty and quality scores; Bernoulli models for convergence. \\
\addlinespace[3pt]
Reward $R$ & Quality-score increase over the retained space when a candidate is accepted; zero otherwise. \\
\addlinespace[3pt]
Discount $\gamma$ & Weight of the next calculation's reward relative to the current reward. \\
\midrule
Belief $b_t$ & Posterior over the unknown quantities, updated from completed calculations and used to select the next action. \\
\bottomrule
\end{tabularx}
\end{table}

At step $t$, the state $s_t=(\xi_t,\boldsymbol\theta_t)$ comprises the known conditions $\xi_t$ and unknown quantities $\boldsymbol\theta_t$ listed in Table~\ref{tab:pomdp}. The \emph{belief} $b_t(\boldsymbol\theta)$ is the posterior distribution of these unknown quantities given the completed calculations. Within the current orbital context, action $a_t$ produces an observation $o_{t+1}\in\Omega$, which updates the belief by\cite{Kaelbling1998POMDP}
\begin{equation}
 b_{t+1}(\boldsymbol\theta)\propto
 O(o_{t+1}\mid\boldsymbol\theta,a_t)\,b_t(\boldsymbol\theta).
 \label{eq:bayes_filter}
\end{equation}
Here, $O(o_{t+1}\mid\boldsymbol\theta,a_t)$ is the likelihood of the observed result given the unknown quantities and action. The proportionality constant normalizes the posterior. Diagnostic results update orbital statistics. Valid candidate scores update the predicted quality of untested spaces through the \emph{input--response} model. QUBO candidates are generated from the latest diagnostic posterior.

\begin{figure}[htpb!]
\centering
\includegraphics[width=227.5bp]{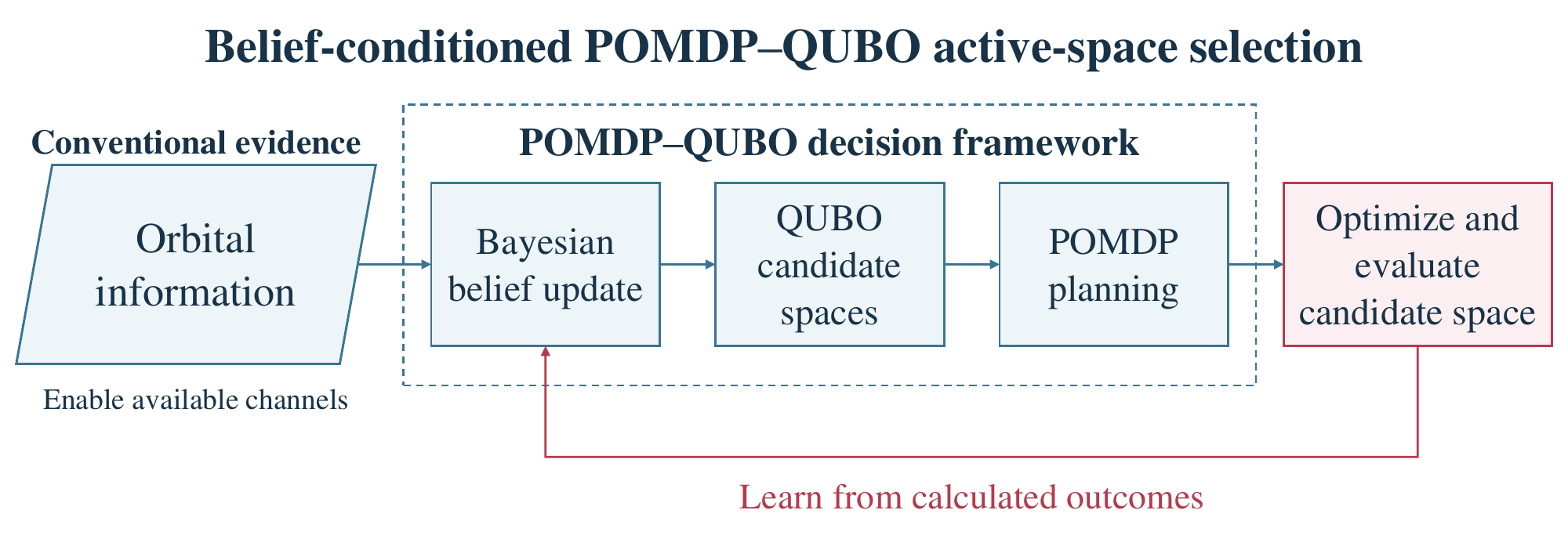}
\caption{Unified POMDP--QUBO framework for active-space selection. Orbital information and Bayesian updates (Sec. \ref{sec:posterior}) inform QUBO candidate construction (Sec. \ref{sec:qubo}) and POMDP planning, while calculated outcomes guide subsequent choices. (Sec. \ref{sec:cycle})}
\label{fig:workflow}
\end{figure}

Figure~\ref{fig:workflow} shows the iterative cycle of belief updating via the {\emph{input--response}} model and calculation planning. A diagnostic first characterizes an orbital window and updates the belief. QUBO then constructs candidate spaces at fixed electron and orbital counts, and the POMDP uses two-step lookahead to select the candidate calculation with the largest expected accepted quality gain. A valid result replaces the retained space when its quality improvement exceeds the threshold; otherwise, the current space is kept. All valid candidate scores update the belief, while failed calculations update only the convergence probability. If an accepted result changes the orbital basis, a new diagnostic is performed before selection continues. The cycle ends when the budget for candidate calculations is exhausted or a stopping criterion is met, returning the retained space.

\subsection{Orbital information and Bayesian updates}\label{sec:posterior}

Three criteria below describe different aspects of an active space, and they are used to update the \emph{belief} $b_t(\boldsymbol\theta)$.

\textbf{Occupation.} For the spin-summed one-particle density matrix $\mathbf D_r$ restricted to $A$, we use
\begin{equation}
 U(A)=\frac{1}{K}\sum_r w_r\left[2\operatorname{Tr}\mathbf D_r-\operatorname{Tr}(\mathbf D_r^2)\right].
 \label{eq:occupation_evidence}
\end{equation}
where $\operatorname{Tr}$ denotes the trace. Each natural occupation $n$ contributes $n(2-n)$, which is zero for empty and doubly occupied orbitals.\cite{Takatsuka1978OddElectrons} Keeping the full density matrix includes the off-diagonal terms when the input orbitals are not natural orbitals.

\textbf{Orbital composition.} Projections onto chosen atomic-valence functions, collectively called the Target, measure the chemical character of a space.\cite{Sayfutyarova2017AVAS} We combine the mean projection with a measure of whether complementary valence directions are both represented:
\begin{equation}
 C(A)=\frac{\alpha}{K}\sum_{i\in A}a_i
 +(1-\alpha)\sum_k\omega_k\det\mathbf G_{k,A}.
 \label{eq:composition_evidence}
\end{equation}
Here $a_i$ is the squared projection of orbital $i$ onto the Target. For each pair of complementary directions $k$, the $2\times K$ matrix $\mathbf Z_{k,A}$ contains their projection amplitudes onto the selected orbitals. Their Gram matrix is $\mathbf G_{k,A}=\mathbf Z_{k,A}\mathbf Z_{k,A}^{\dagger}$, where $\dagger$ denotes conjugate transpose. Its determinant is small when only one direction is represented and increases when both are included. The nonnegative weights $\omega_k$ sum to one, and $0\leq\alpha\leq1$ sets the balance between mean projection and complementary coverage.

\textbf{Entanglement.} Single-orbital entropy measures the correlation of an orbital with the rest of the system, while mutual information measures the correlation between two orbitals:\cite{Legeza2003Entropy,Rissler2006OrbitalMI,Boguslawski2012Entanglement,Luo2017EDGA}
\begin{equation}
 \begin{aligned}
 s_{ir}&=-\operatorname{Tr}(\rho_{ir}\ln\rho_{ir}),\\
 s_{ijr}&=-\operatorname{Tr}(\rho_{ijr}\ln\rho_{ijr}),\\
 I_{ijr}&=\tfrac12(s_{ir}+s_{jr}-s_{ijr}),\qquad i\ne j.
 \end{aligned}
 \label{eq:entanglement_evidence}
\end{equation}
Here $\rho_{ir}$ and $\rho_{ijr}$ are the one- and two-orbital reduced density matrices for electronic state $r$, obtained by tracing out the other orbitals.\cite{Boguslawski2012Entanglement, Baiardi2020DMRGReview, Bensberg2023CorrespondingActiveSpaces} Their \textit{von} Neumann entropies are $s_{ir}$ and $s_{ijr}$, respectively; $\ln$ denotes the natural logarithm and $I_{iir}=0$. For active-space selection, we define the score $J(A)$ as an equal-weight combination of normalized single-orbital entropies and internal mutual information, averaged with the state weights $w_r$ (Supporting Information).

The active-space quality score $Q(A)$ combines the criteria above with equal weights: $Q(A)=[U(A)+C(A)]/2$ for the \textbf{\emph{2-channel}} scheme and $Q(A)=[U(A)+C(A)+J(A)]/3$ for the \textbf{\emph{3-channel}} scheme. Before a candidate is calculated, its quality is uncertain. The belief $b_t(\boldsymbol\theta)$ defined in Section~\ref{sec:pomdp}, together with the joint \emph{input--response} model below, gives the predictive distribution of this score. The POMDP uses that distribution to evaluate expected quality gains. The reward is the calculated quality increase relative to the retained space when a candidate is accepted, and zero otherwise.

An \emph{input} model maintains Gaussian posteriors for occupation and, when available, entanglement statistics in the diagnostic window. Composition descriptors are computed directly from the orbitals. A \emph{response} model predicts candidate component scores, combining input uncertainty with residual covariance for differences such as orbital relaxation. Shared orbital information correlates candidate predictions. Gaussian Bayesian updates implement Eq.~\ref{eq:bayes_filter} for these statistics and scores.\cite{Kalman1960Filtering, Sarkka2023BayesianFiltering} Every valid candidate score updates the quality predictions used by the POMDP planner, whether the space is accepted or rejected. QUBO coefficients use the latest diagnostic posterior. Convergence is described separately by a Beta--Bernoulli model.\cite{Bishop2006PRML}
Details are given in the Supporting Information.





\subsection{Constructing candidates with QUBO}\label{sec:qubo}

Let $x_i=1$ if orbital $i$ in the diagnostic window $W$ is selected, and zero otherwise. The conditions $\sum_i x_i=K$ and $\sum_i\nu_i x_i=N_e$ fix the orbital and electron counts, with reference occupations $\nu_i\in\{0,1,2\}$. All singly occupied reference orbitals remain active. These integer reference occupations enforce the electron-count constraint. The QUBO occupation terms use posterior means of statistics derived from the diagnostic one-particle density matrices (Section~\ref{sec:posterior}).

\begin{figure}[htbp!]
\centering
\includegraphics[scale=0.275,bb=75 0 1000 500]{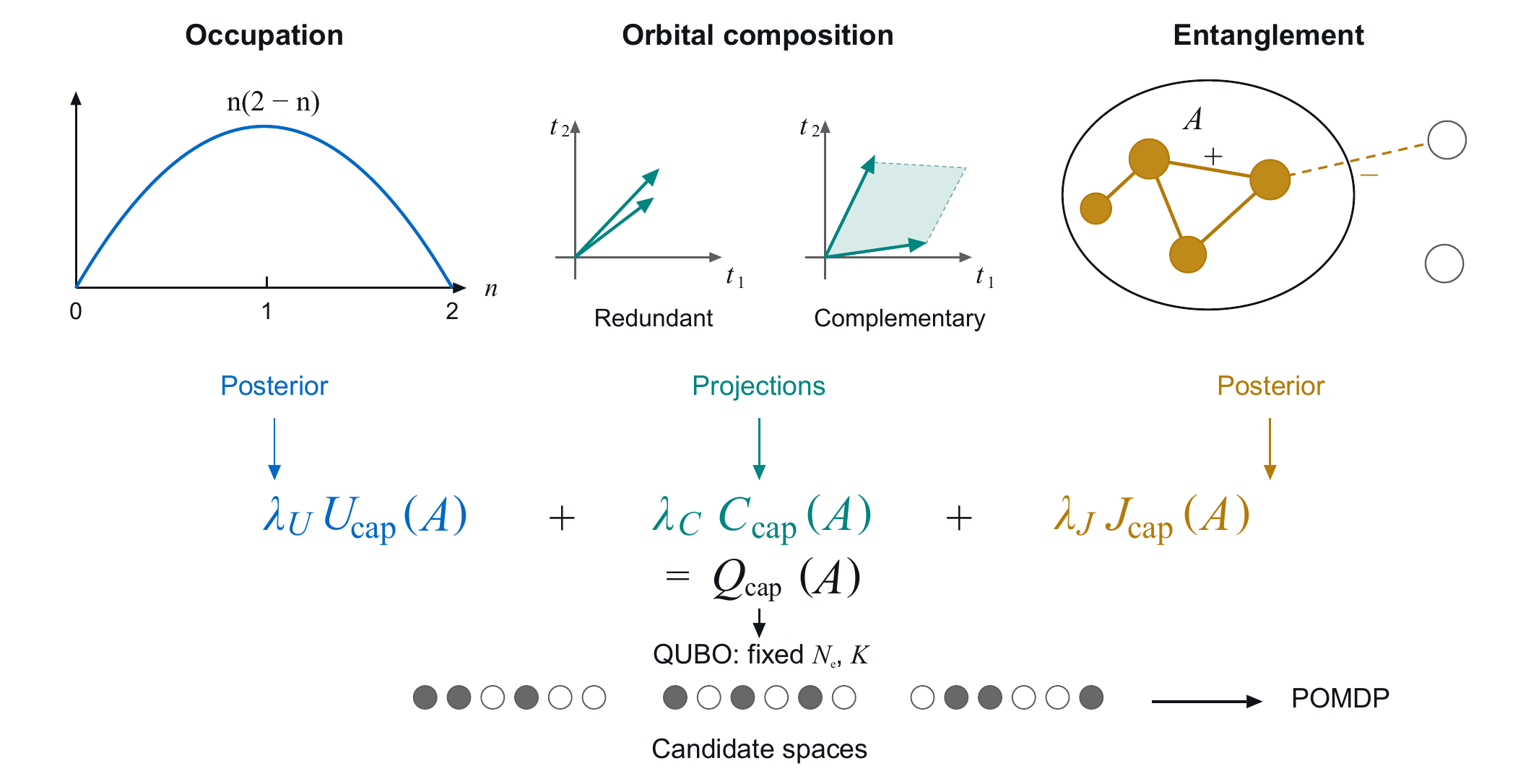}
\caption{Construction of the QUBO capture score from three evidence channels. Occupation and entanglement contributions use input posterior statistics; composition uses target projections. Green vectors represent orbital projections onto target directions $t_1$ and $t_2$: nearly parallel projections provide redundant coverage, whereas independent directions complement one another. Internal mutual-information links contribute positively, while links crossing the selected-space boundary are penalized. The combined score supplies candidate spaces for POMDP selection at fixed electron and orbital counts.}
\label{fig:qubo_evidence}
\end{figure}

As shown in Figure~\ref{fig:qubo_evidence}, each of the three criteria expands naturally into single-orbital and orbital-pair terms. Occupation contributes density terms, composition contributes projections and complementary coverage, and entanglement contributes orbital entropies and internal or boundary correlations. Their combined capture score is
\begin{equation}
 \begin{aligned}
 Q_{\mathrm{cap}}(\boldsymbol x)
 &=\sum_{a\in\mathcal E}\lambda_a Q_{a,\mathrm{cap}}(\boldsymbol x)\\
 &=\sum_i \ell_i x_i+\sum_{i<j}q_{ij}x_ix_j.
 \end{aligned}
 \label{eq:capture}
\end{equation}
Here $\mathcal E\subseteq\{U,C,J\}$ is the set of selected criteria, and $Q_{a,\mathrm{cap}}$ is the contribution from component $a$. $Q_{\mathrm{cap}}$ and $Q$ use the same fixed, equal weights $\lambda_a$: $1$ for a single criterion, $1/2$ for two criteria, and $1/3$ for three. The linear coefficient $\ell_i$ describes one orbital, and $q_{ij}$ describes an unordered orbital pair. These coefficients use posterior means for occupation and entanglement, and direct projections for composition. Occupation and composition have the same functional forms in $Q_{\mathrm{cap}}$ and $Q$. The quantum information theory (QIT) capture term uses window-normalized information and penalizes correlations across the selection boundary; the calculated-space score uses normalized entropies and mutual information within the selected space.

Maximizing $Q_{\mathrm{cap}}$ is equivalent to minimizing its negative as a QUBO problem.\cite{Lucas2014Ising,Kochenberger2014UBQPSurvey,Glover2022QUBOTutorial} We enumerate all feasible orbital combinations when their number is at most 100\,000. For larger domains, simulated annealing uses 64 random feasible starting spaces and exchanges orbitals while preserving the electron and orbital counts.\cite{Kirkpatrick1983SimulatedAnnealing} We deduplicate the combinations scored during the search, exclude the retained space and spaces already attempted in the current orbital basis, and rank the remainder by $Q_{\mathrm{cap}}$. The candidate pool $\mathcal C_t$ retains the 15 highest-scoring eligible spaces, or all eligible spaces if fewer are found. This preset pool-size limit bounds the number of candidate pairs compared in two-step POMDP planning. The planner then selects the next candidate to calculate. Full search settings are given in the Supporting Information.

\subsection{Testing and retaining a candidate}\label{sec:cycle}

A candidate replaces the retained space $A_0$ when its calculated quality satisfies $Q_{\mathrm{out}}>Q_0+\varepsilon$, where $Q_0=Q(A_0)$ and $\varepsilon=10^{-6}$ in our implementation. The reward is the accepted gain $R_t=Q_{\mathrm{out}}-Q_0$; other outcomes have zero reward. To choose the next calculation, we use finite-horizon POMDP planning\cite{Smallwood1973FiniteHorizon,Kaelbling1998POMDP} over at most two successive candidates:
\begin{equation}
 \begin{aligned}
 V_2(A;b_t)&=\mathbb E[R_t+\gamma R_{t+1}\mid b_t,A],\\
 A_t^*&=\argmax_{A\in\mathcal C_t}V_2(A;b_t),\qquad \gamma=1.
 \end{aligned}
 \label{eq:policy}
\end{equation}
The expectation includes the possible quality gains, numerical convergence, and what the first result would teach us before choosing the second candidate. An accepted first result ends that simulated sequence. Only the first calculation is carried out, after which the predictions and choice are updated.

A converged result updates the candidate-quality posterior whether or not the candidate is accepted. Acceptance changes the retained space; when the orbitals also change, a new diagnostic characterizes them before selection continues. The procedure returns the retained space when the stopping criterion is met or the available calculations are used.

\section{Computational Details}\label{sec:protocol}

The core POMDP–QUBO active-space selection engine is implemented in {\sc Python} and released as open-source software. The current interface supports coupling with {\sc PySCF},\cite{Sun2018PySCF,Sun2020PySCF} {\sc OpenMolcas},\cite{FdezGalvan2019OpenMolcas} {\sc Block2},\cite{Zhai2023Block2} {\sc QCMaquis},\cite{Keller2015MPO} and {\sc Kylin}.\cite{Xie2023Kylin} It is also integrated as a stand-alone module within the development version of the {\sc Kylin} package.

The evaluated systems and basis sets are summarized in Table~\ref{tab:systems}. The small-molecule tests use 6-31G*,\cite{Hehre1972SplitValence,Hariharan1973Polarization} and 1,2-dioxetanone uses ANO-RCC-VDZP.\cite{Roos2004ANORCC}
Active-space selection proceeds as follows. For the \emph{\textbf{small molecules}} the target spaces are defined by atomic-valence projections; the procedure recovers the expected valence orbitals (bonding/antibonding pairs, main-group valence sets, and the open-shell $\pi/\pi^*$ space of triplet O$_2$). 
For \textbf{\emph{TiF$_3$}} the five Ti 3d orbitals are retained as seeds. Selection is carried out in a fixed-orbital CAS(33e,25o) DMRG calculation whose target includes Ti 3s/3p/3d/4s/4p and F 2s/2p character; the resulting spaces are compared with published reference excitation energies.\cite{Sayfutyarova2018EPRDMRG}
For \emph{\textbf{1,2-dioxetanone}} the target comprises C 2s/2p and O 2p functions. Selection is performed at the CAS(16e,13o) level with two equally weighted singlet states; a diagnostic window is constructed from the retained active orbitals together with additional orbitals of appreciable target character or near-frontier energy, and successive candidates are evaluated with DMRG (preserved state $M=1000$).

\begin{table}[htbp!]
\centering
\caption{Evaluated systems and their computational details.}
\label{tab:systems}
\scriptsize
	\begin{tabular}{cccc}
		\hline
		\hline
System & & Basis set & Remarks  \\
		\hline   
 Small molecules  & & 6-31G*   &   H$_2$, HF, H$_2$O, NH$_3$, triplet O$_2$  \\   
 		\hline               
 \multirow{2}*{1,2-Dioxetanone} & &ANO-RCC-VDZP    & Cl${_{OO}}$ $^a$, Cl${_{CC}}$ $^a$\\  
                & & + DKH2 $^b$  & \smin{} $^a$ \\  
                		\hline
 \multirow{2}*{TiF$_3$ }        & & ANO-RCC        &  Ti: [7s6p5d3f2g1h], F: [4s3p2d1f]\\
                & & + DKH2 $^b$  &  planar $D_{3h}$ geometry \\
       \hline
       \hline
	\end{tabular}
   \begin{tablenotes}
      \scriptsize
      \item {a : Optimized conical intersection / \smin{} structures from Ref. \citenum{Ma2020FullValenceDioxetanone}}
      \item {b : All-electron basis sets of atomic natural orbital type developed for relativistic calculations,\cite{Roos2004ANORCC} together with 2nd-order Douglas-Kroll-Hess Hamiltonian\cite{Hess1986DKH}  }
    \end{tablenotes}
\end{table}

\section{Results and Discussion}\label{sec:results}
\setcounter{dbltopnumber}{3}

\subsection{Valence spaces in small molecules}

\begin{table*}[htbp!]
\centering
\caption{Selected valence spaces and orbital isosurfaces in the small-molecule tests.}
\label{tab:small_main}
\small
\setlength{\tabcolsep}{3pt}
\renewcommand{\arraystretch}{1.0}
\begin{tabular}{@{}m{0.12\textwidth}>{\centering\arraybackslash}m{0.65\textwidth}>{\raggedright\arraybackslash}m{0.19\textwidth}@{}}
\toprule
Molecule / space & Selected valence orbitals & Occupation / symmetry \\
\midrule
\begin{tabular}[c]{@{}l@{}}H$_2$\\[-1pt]{\scriptsize CAS(2e,2o)}\\[-1pt]{\scriptsize $P_\Sigma=2.0000$}\end{tabular} & \includegraphics[width=0.285714\linewidth]{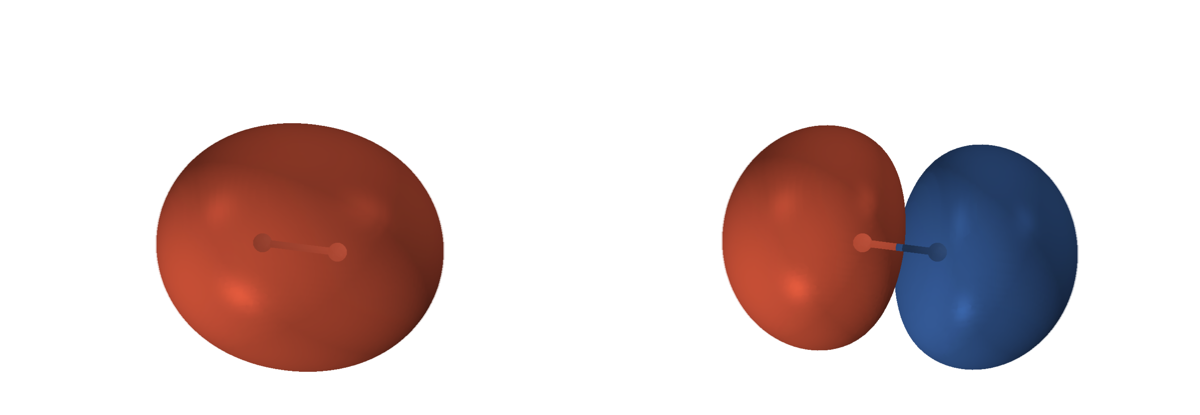} & \begin{minipage}{\linewidth}\footnotesize $\nu=(2,0)$\\[2pt]{$\sigma_g/\sigma_u^*$ pair}\end{minipage} \\
\addlinespace[1pt]
\begin{tabular}[c]{@{}l@{}}HF\\[-1pt]{\scriptsize CAS(8e,5o)}\\[-1pt]{\scriptsize $P_\Sigma=4.9949$}\end{tabular} & \includegraphics[width=0.714286\linewidth]{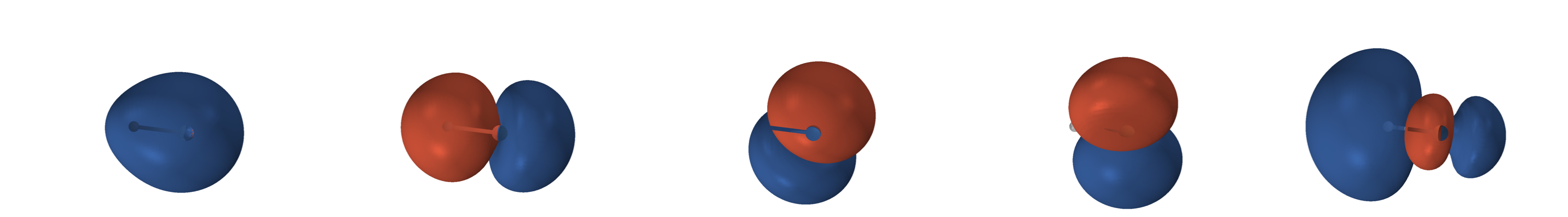} & \begin{minipage}{\linewidth}\footnotesize $\nu=(2,2,2,2,0)$\\[2pt]{two $\pi$ orbitals}\end{minipage} \\
\addlinespace[1pt]
\begin{tabular}[c]{@{}l@{}}H$_2$O\\[-1pt]{\scriptsize CAS(8e,6o)}\\[-1pt]{\scriptsize $P_\Sigma=5.9887$}\end{tabular} & \includegraphics[width=0.857143\linewidth]{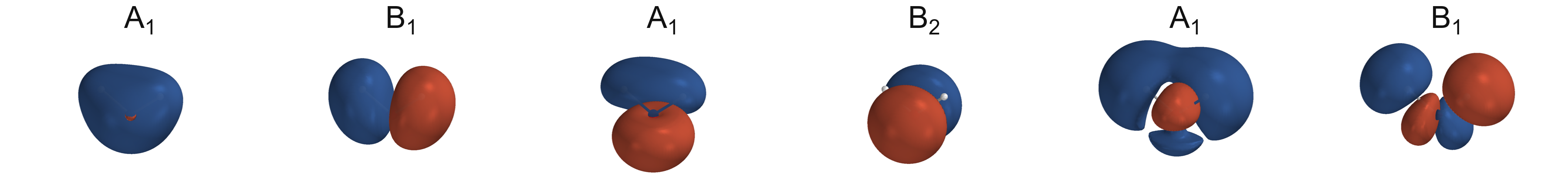} & \begin{minipage}{\linewidth}\footnotesize $\nu=(2,2,2,2,0,0)$\\[2pt]{$C_1\leftrightarrow C_{2v}$\\All six $\sigma_i=1.0000$}\end{minipage} \\
\addlinespace[1pt]
\begin{tabular}[c]{@{}l@{}}NH$_3$\\[-1pt]{\scriptsize CAS(8e,7o)}\\[-1pt]{\scriptsize $P_\Sigma=6.9775$}\end{tabular} & \includegraphics[width=1.000000\linewidth]{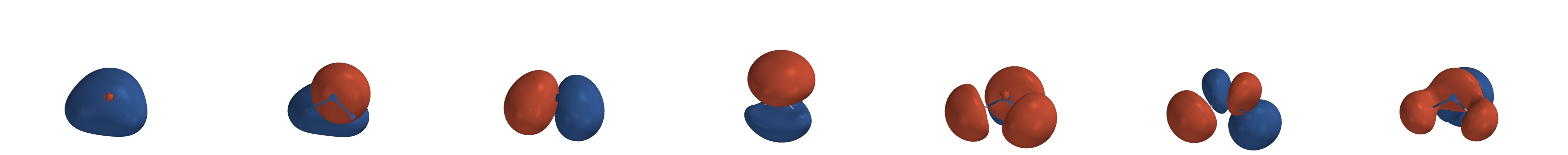} & \begin{minipage}{\linewidth}\footnotesize $\nu=(2,2,2,2,0,0,0)$\\[2pt]{full valence set}\end{minipage} \\
\addlinespace[1pt]
\begin{tabular}[c]{@{}l@{}}O$_2$\\[-1pt]{\scriptsize CAS(6e,4o)}\\[-1pt]{\scriptsize $P_\Sigma=4.0000$}\end{tabular} & \includegraphics[width=0.571429\linewidth]{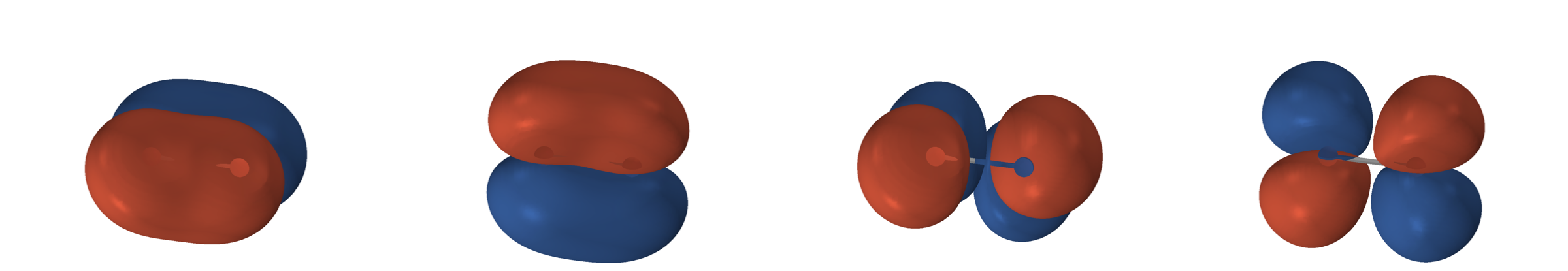} & \begin{minipage}{\linewidth}\footnotesize $\nu=(2,2,1,1)$\\[2pt]{complete $\pi/\pi^*$ pairs}\end{minipage} \\
\bottomrule
\end{tabular}
\par\vspace{3pt}
\begin{minipage}{0.98\textwidth}\footnotesize For all five molecules, the initial and final retained orbital sets are identical. $\nu$ lists RHF/ROHF reference occupations in the plotted order. 
Selection uses $C_1$ symmetry; H$_2$O is also checked in $C_{2v}$ and shown with its symmetry labels using $xz$ molecular plane.\end{minipage}
\end{table*}

As an appetizer, Table~\ref{tab:small_main} shows POMDP-QUBO framework can easily recover the valence active orbitals of all five molecules. The restricted Hartree-Fock (RHF) orbitals are used as the initial orbitals, the present 2-channel scheme is equivalent to the AVAS orbital selection approach (i.e. $C(A)$ channel works), since either 2 or 0 orbital occupations for RHF anastz, the $U(A)$ channel is not take effect.

H$_2$ gives the $\sigma_g/\sigma_u^*$ bonding-antibonding pair. HF, H$_2$O, and NH$_3$ retain occupied bonding and non-bonding orbitals together with one, two, and three empty valence orbitals, respectively. For each space, the total target projection $P_\Sigma$ is close to the active-orbital count $K$. In triplet O$_2$, selection preserves both components of the bonding $\pi$ and anti-bonding $\pi^*$ pairs; the two singly occupied $\pi^*$ orbitals retain the open-shell configuration.

The H$_2$O comparison tests whether imposing point-group symmetry changes the selected space. The $C_1$ and $C_{2v}$ calculations give the same six-dimensional subspace. All six overlap singular values round to 1.0000, and both RHF energies round to $-76.009132\ E_{\mathrm h}$. The table shows the equivalent $C_{2v}$ orbitals and their symmetry labels. Thus, the comparison checks the whole selected subspace, allowing individual orbitals to differ by rotations within it.\cite{Bjorck1973PrincipalAngles}

\subsection{Spectral calculation of metal compound \texorpdfstring{TiF$_3$}{TiF3}}

TiF$_3$ tests fixed-orbital selection for several low-lying states of an open-shell transition-metal molecule. The five lowest doublet roots form a ground state and two nearly degenerate excited pairs. 
Both the 2- and 3-channel POMDP-QUBO implementation can obtained the desired active space. 
Table~\ref{tab:tif3_main} compares the initial and final vertical excitation energies with the literature values. At the same CAS(33e,25o) size and $M=1000$, both pair-averaged excitation energies move closer to the published values.

\begin{table}[htbp!]
\centering
\caption{Pair-averaged TiF$_3$ excitation energies and absolute deviations from the literature, in cm$^{-1}$.}
\label{tab:tif3_main}
\small
\renewcommand{\arraystretch}{1.15}
\setlength{\tabcolsep}{2pt}
\begin{tabular*}{\columnwidth}{@{\extracolsep{\fill}}lrrrc@{}}
\toprule
State & Initial & Final & Ref.~\citenum{Sayfutyarova2018EPRDMRG} & Abs. deviation \\
 & & & & Initial $\to$ final \\
\midrule
$1\,{}^2E''$ & 5255 & 5733 & 5785 & $530\to52$ \\
$1\,{}^2E'$ & 22061 & 22331 & 22538 & $477\to207$ \\
\bottomrule
\end{tabular*}
\end{table}

\begin{figure*}[htbp!]
\centering
\includegraphics[width=\textwidth]{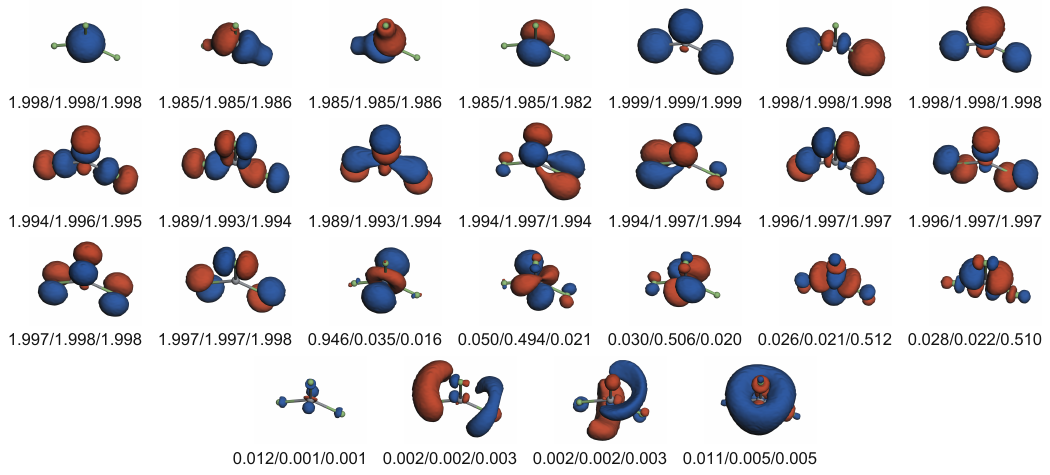}
\caption{The final 25 active orbitals of TiF$_3$. Values below each orbital are $n({}^2A_1')/n(1\,{}^2E'')/n(1\,{}^2E')$ in the displayed orbital basis; each $E$-state value averages the two corresponding roots.} 
\label{fig:tif3_active_orbitals}
\end{figure*}

The 5 retained Ti~3d seed orbitals appear at the right of the third row in Figure~\ref{fig:tif3_active_orbitals}. The predominantly Ti~$3d_{z^2}$ orbital has a ground-state occupation of 0.946. The lower and upper excited-state pairs populate two distinct pairs of Ti~3d orbitals, with pair-averaged occupations of about 0.5 per orbital. This pattern agrees with the $^2A_1'$ ground state and metal-centered $1\,{}^2E''$ and $1\,{}^2E'$ ligand-field excitations described by Sayfutyarova and Chan.\cite{Sayfutyarova2018EPRDMRG} Sixteen orbitals remain nearly doubly occupied; the four weakly occupied correlation orbitals appear in the bottom row. The literature likewise extends the five-orbital space with ligand and Ti~s/p orbitals to describe core-valence and valence-virtual correlation.\cite{Sayfutyarova2018EPRDMRG}

The refinement replaces a predominantly Ti~$p_z$ orbital with an orbital combining Ti~s/$d_{z^2}$ character and amplitude on the F ligands (rightmost orbital in the bottom row). This supplies an additional metal-ligand correlation orbital around the retained $3d_{z^2}$ direction. Its occupation is 0.011 in the ground state and about 0.005 in each excited pair. The substitution lowers the ground-state energy by 7.007~m$E_{\mathrm h}$ and the two excited-pair averages by 4.828 and 5.777~m$E_{\mathrm h}$. The greater ground-state stabilization raises the excitation energies by 478 and 270~cm$^{-1}$, reducing their absolute deviations from 530 to 52 and from 477 to 207~cm$^{-1}$. Spatial symmetry is not imposed here; the final within-pair splittings are 3.7 and 3.2~cm$^{-1}$.

\subsection{Refined active-space selection in 1,2-dioxetanone}

\begin{figure}[htbp!]
\centering
\includegraphics[width=\columnwidth]{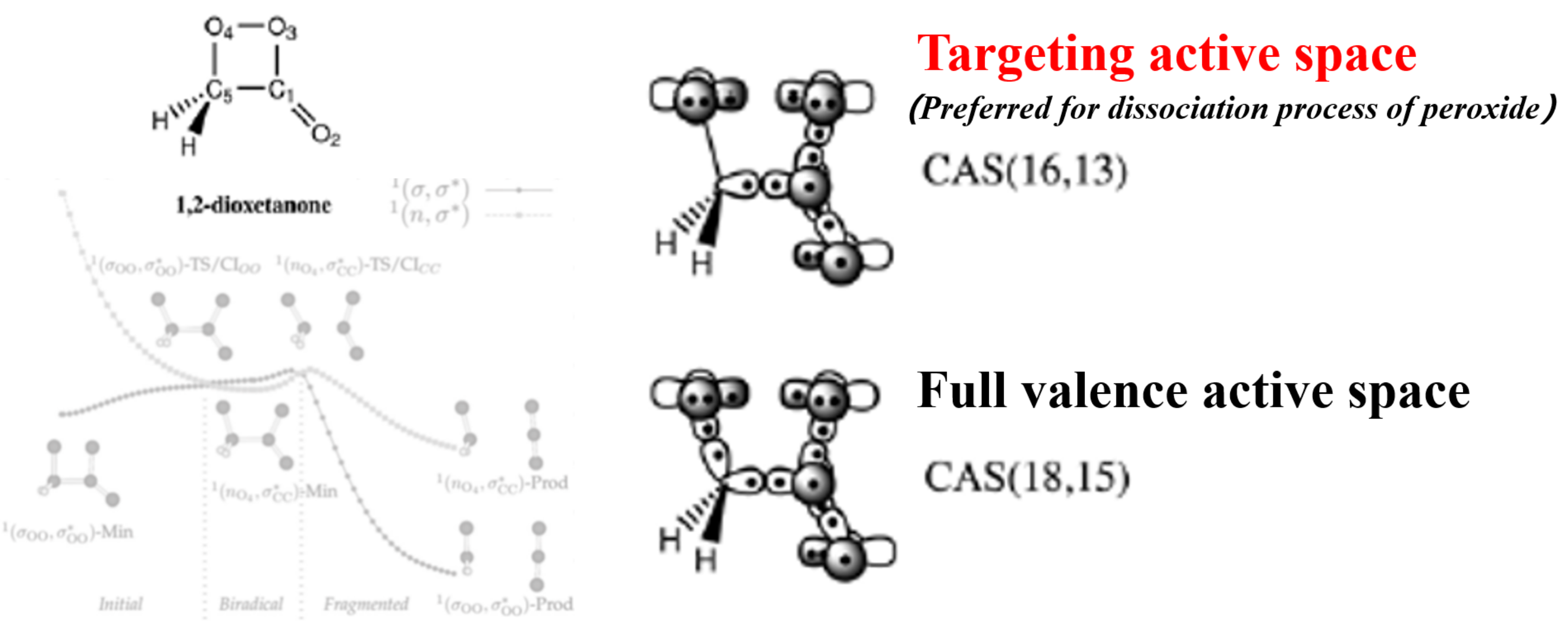}
\caption{Targeted CAS(16e,13o) selection for 1,2-dioxetanone, compared with the larger CAS(18e,15o) valence space.}
\label{fig:targeted_dioxetanone}
\end{figure}

Figure~\ref{fig:targeted_dioxetanone} contrasts the targeted CAS(16e,13o) with the larger CAS(18e,15o) valence space for 1,2-dioxetanone. Here we test whether selection at this fixed size reproduces the orbital spaces established in the literature. We examine three geometries: the \smin{} and the two conical intersections, \cioo\ and \cicc.\cite{Ma2020FullValenceDioxetanone}

\begin{figure*}[htbp!]
\centering
\includegraphics[width=\textwidth]{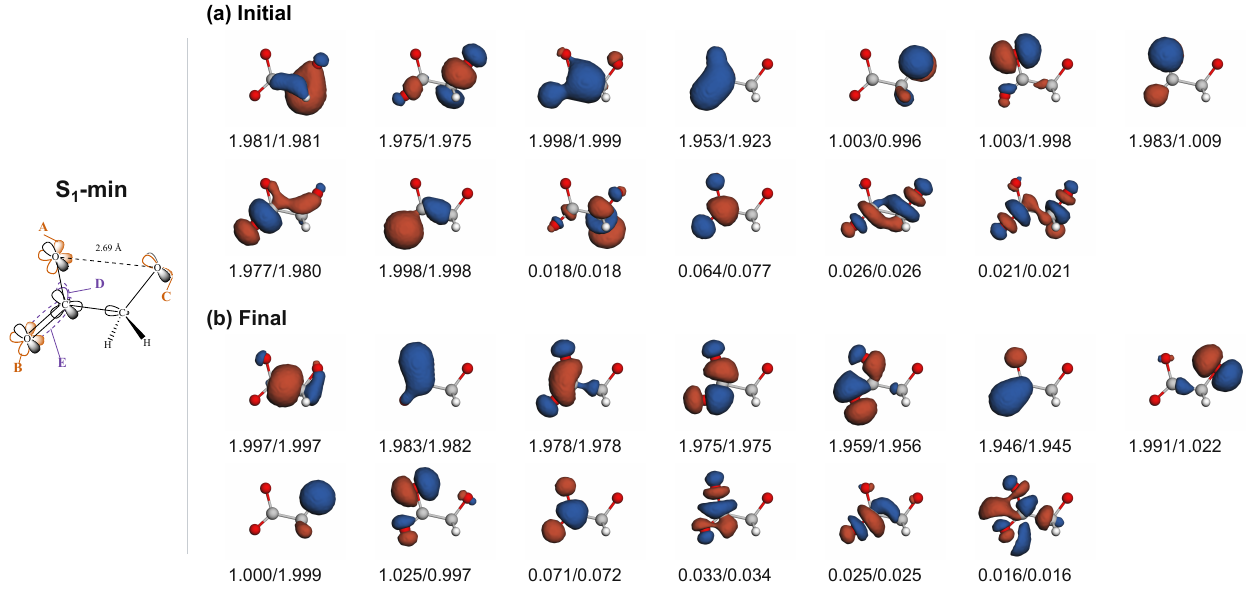}
\caption{All 13 active orbitals at \smin{} before (a) and after (b) 3-channel POMDP-QUBO selection, the main orbital changes are also shown in the schematic on the left.
Values below each orbital are orbital occupation numbers for $n(S_0)/n(S_1)$ states.
}
\label{fig:nx1_active_orbitals}
\end{figure*}

\begin{figure*}[tp]
\centering
\includegraphics[width=\textwidth]{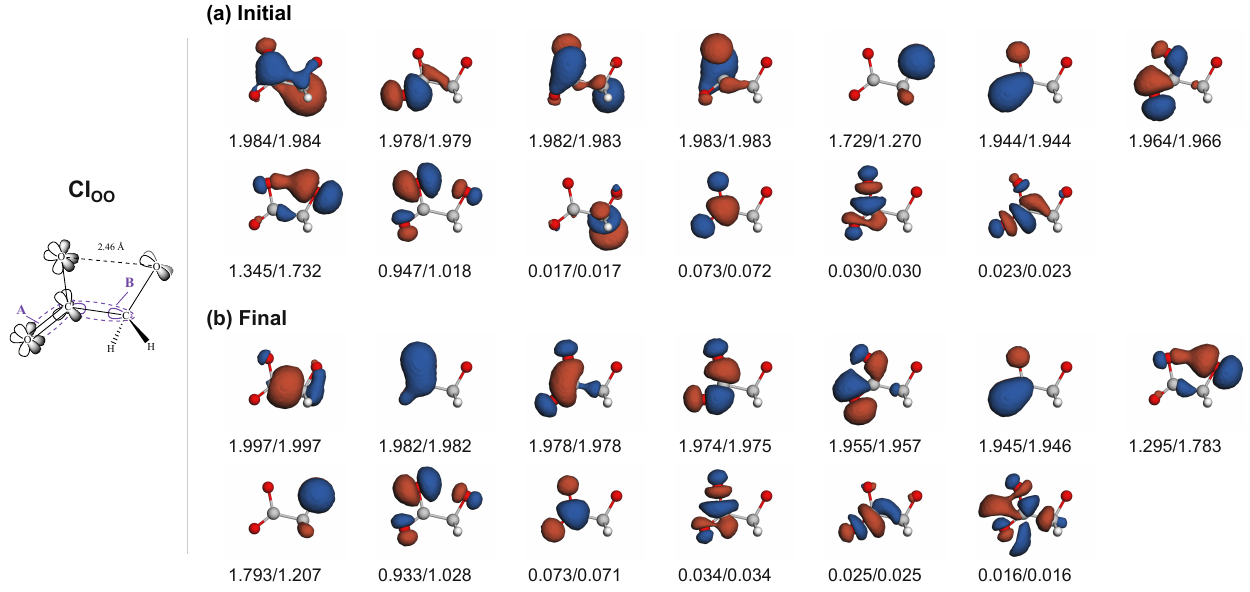}
\caption{All 13 active orbitals at \cioo\ before (a) and after (b) 3-channel POMDP-QUBO selection. 
Other conventions follow Figure~\ref{fig:nx1_active_orbitals}.}
\label{fig:cioo_active_orbitals}
\end{figure*}

We assess orbital consistency through the singular values of the AO-metric overlap between the selected and reference spaces.\cite{Bjorck1973PrincipalAngles} This comparison is invariant to orbital rotations within either space. At all three geometries, the final space recovers all 13 reference directions, and all 13 singular values round to 1.0000 (Table~\ref{tab:results}). The selected spaces therefore match the respective literature reference subspaces to the reported precision.

\begin{table}[htbp]
\centering
\caption{Agreement with the literature CAS(16e,13o) reference spaces. Coverage counts overlap singular values $\geq0.95$; $\sigma_{\min}$ refers to the final space.}
\label{tab:results}
\small
\renewcommand{\arraystretch}{1.15}
\setlength{\tabcolsep}{3pt}
\begin{tabular*}{\columnwidth}{@{\extracolsep{\fill}}lccc@{}}
\toprule
 & \multicolumn{2}{c}{Reference coverage} & \\
\cmidrule(lr){2-3}
Geometry & Initial & Final & $\sigma_{\min}$ \\
\midrule
\smin{} & 8/13 & 13/13 & 1.0000 \\
\cioo & 11/13 & 13/13 & 1.0000 \\
\cicc & 13/13 & 13/13 & 1.0000 \\
\bottomrule
\end{tabular*}
\end{table}

\begin{figure*}[!t]
\centering
\includegraphics[width=0.97\textwidth]{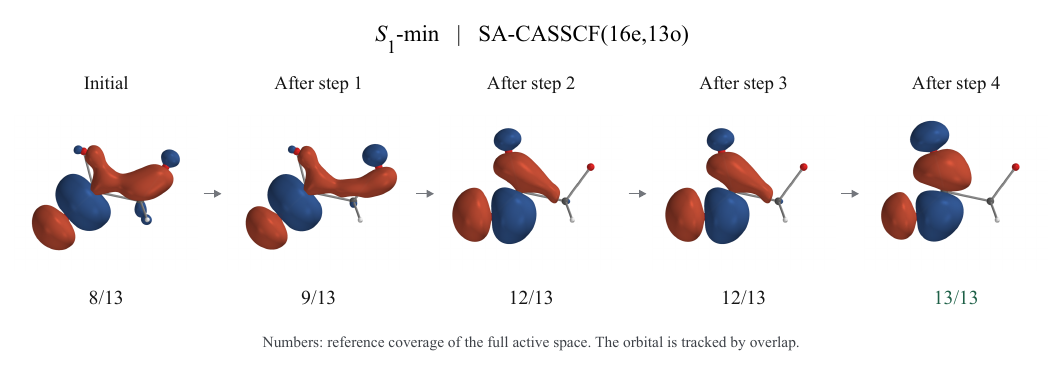}
\caption{Evolution of a representative active orbital at \smin{} through four accepted SA-CASSCF optimizations. Orbitals are matched by overlap and phase-aligned; isosurfaces are $\pm0.05$ a.u. Fractions give reference coverage of the full active space.}
\label{fig:smin_cas_orbital_evolution}
\end{figure*}

\begin{table*}[t]
\centering
\caption{Selection effort and final retained spaces at \smin{} and \cioo. $N_{\mathrm{opt}}$ counts candidate optimizations; coverage is initial $\to$ final. $\Delta E_{21}=E_2-E_1$.}
\label{tab:cas_nc_main}
\small
\renewcommand{\arraystretch}{1.15}
\setlength{\tabcolsep}{4pt}
\begin{tabular*}{\textwidth}{@{\extracolsep{\fill}}llcc c S[table-format=-3.6] S[table-format=1.5]@{}}
\toprule
Geometry & Solver & Channels & $N_{\mathrm{opt}}$ & Coverage & {$E_{\mathrm{av}}/E_{\mathrm h}$} & {$\Delta E_{21}$/eV} \\
\midrule
\smin{} & DMRG & $U,C,J$ & 2 & $8/13\to13/13$ & -301.831972 & 0.17230 \\
 & CASSCF & $U,C$ & 4 & $8/13\to13/13$ & -301.831973 & 0.17231 \\
\addlinespace[3pt]
\cioo & DMRG & $U,C,J$ & 1 & $11/13\to13/13$ & -301.831519 & 0.04341 \\
 & CASSCF & $U,C$ & 6 & $11/13\to12/13$ & -301.823384 & 0.63151 \\
\bottomrule
\end{tabular*}
\end{table*}

At \smin{}, three-channel DMRG refinement increases reference coverage from 8/13 to 11/13 and then 13/13 in two candidate optimizations. At \cioo, one candidate optimization increases coverage from 11/13 to 13/13. At \cicc, the initial CASSCF space already matches the reference and is retained. At this near-dissociation geometry, the key fractionally occupied orbitals lie in the frontier energy region, facilitating their inclusion in the initial active space. Thus, the same fixed-size selection framework recovers or retains the reference orbital space at each of the three geometries.

Figures~\ref{fig:nx1_active_orbitals} and~\ref{fig:cioo_active_orbitals} show the complete initial and final orbital sets at \smin{} and \cioo. These changes refine the orbital content while keeping 16 electrons in 13 active orbitals. Energy comparisons are given in the Supporting Information.

\subsection{2-channel \textit{vs.} 3-channel}\label{sec:cas_nc}


For dioxetanone, we compare 2-channel CASSCF ($U,C$) with 3-channel DMRG ($U,C,J$), then examine the contribution of response learning within the 2-channel route.

The 2-channel scheme reaches 13/13 at \smin{}, but requires four candidate optimizations, compared with two for the 3-channel route. Its coverage follows 8/13, 9/13, 12/13, 12/13, and 13/13; all four candidates are accepted. The final mean energies of the two routes differ by $2.9\times10^{-7}\ E_{\mathrm h}$, and both state separations are approximately 0.1723~eV (Table~\ref{tab:cas_nc_main}).

During the four POMDP-guided state-averaged CASSCF optimizations at \smin{}, the tracked active orbital becomes increasingly concentrated on the O--C--O unit, with reduced admixture from the CH$_2$O unit (Figure~\ref{fig:smin_cas_orbital_evolution}). L\"owdin analysis shows that the O-C-O contribution increases from 89.2\% to 97.5\%, whereas the state-averaged occupation changes only from 1.978 to 1.975. This evolution illustrates the refinement of nearly doubly occupied valence-orbital character within the fixed CAS(16e,13o), beyond the changes apparent from occupation numbers alone.

At \cioo, the 2-channel scheme reaches 12/13 within its six-optimization budget. All six tested candidates have that coverage; three are accepted, and the search ends on its optimization budget. The retained space has a two-state separation of 0.63151~eV and a mean energy 8.13~m$E_{\mathrm h}$ above the 3-channel DMRG result. Thus, the QIT-enabled DMRG route recovers the full reference space at both geometries with fewer candidate optimizations. Each CAS search uses four passive diagnostics; diagnostic and optimization counts are reported separately in the Supporting Information.

\subsection{Response learning and two-step planning}\label{sec:response_planning}

At the fourth \smin{} CASSCF decision, three candidate choices share the input orbitals, diagnostics, 15-member pool, and remaining budget. The full model selects candidate-A. Retaining response learning but removing the cross-candidate covariances from two-step planning selects candidate-B, which is also selected by one-step planning. Retaining diagnostics and two-step planning but omitting earlier candidate-score updates selects candidate-C. All three candidates are optimized with the state-averaged CASSCF/DMRG-SCF protocol.

\begin{figure*}[!t]
\centering
\includegraphics[width=0.8375\textwidth]{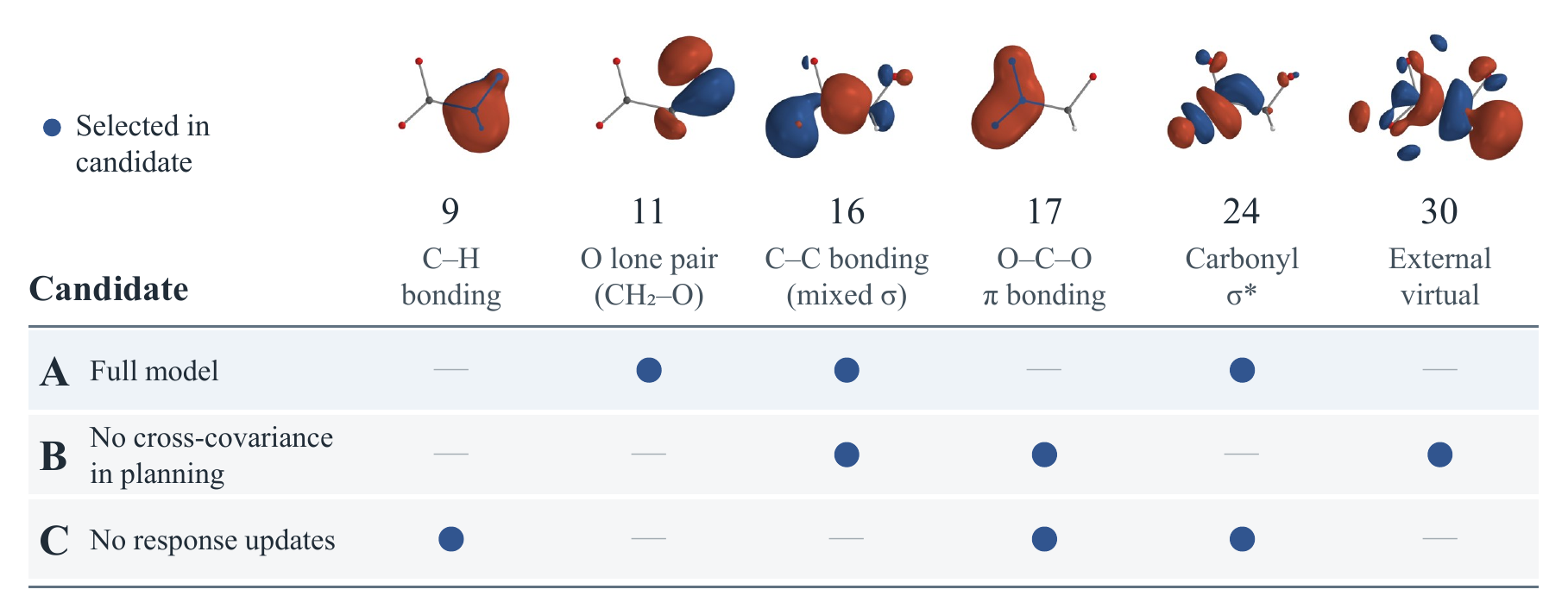}
\caption{A/B/C orbital selections from the same input orbital basis at the fourth \smin{} decision, before candidate optimization. Ten shared orbitals are omitted.}
\label{fig:abc_selection}
\end{figure*}

Figure~\ref{fig:abc_selection} compares the three selections in their common input orbital basis. Candidate-A includes the oxygen lone pair on the CH$_2$O unit (orbital index 11, abbr. as 11), a mixed C--C $\sigma$-bonding orbital (16), and carbonyl $\sigma^*$ (24). Candidate-B omits 11 and 24; Candidate-C omits 11 and underrepresents the C-C direction. Orbitals 30 in candidate-B and 9 in candidate-C have less than 0.2\% projection weight in the reference active space. Candidate-A still lacks the O-C-O $\pi$ direction (17).

\begin{table}[htbp!]
\centering
\caption{CASSCF outcomes of the three \smin{} fourth-step candidates. $\Delta E_{\mathrm{av}}$ is relative to A.}
\label{tab:decision_example}
\small
\renewcommand{\arraystretch}{1.15}
\setlength{\tabcolsep}{3pt}
\begin{tabular*}{\columnwidth}{@{\extracolsep{\fill}}lc S[table-format=1.6] S[table-format=1.3]@{}}
\toprule
Candidate & Coverage & {$Q$} & {$\Delta E_{\mathrm{av}}$/m$E_{\mathrm h}$} \\
\midrule
A & 13/13 & 0.404431 & 0.000 \\
B & 12/13 & 0.399604 & 4.503 \\
C & 12/13 & 0.399603 & 4.503 \\
\bottomrule
\end{tabular*}
\end{table}

Under the full model, candidate-A has a lower immediate expected gain than candidate-B (0.06764 versus 0.06990), but a higher two-step value (0.08613 versus 0.08529). Its calculation recovers all 13 reference directions, whereas candidate-B and candidate-C remain at 12/13 (Table~\ref{tab:decision_example}). The mean energy of candidate-A is lower by 4.50~m$E_{\mathrm h}$, and its quality exceeds both controls by approximately 0.00483. In this decision, accounting for what a calculation can reveal about other candidates leads to the more complete space.

Response updates reduce the mean absolute quality-prediction error for the same \smin{} CASSCF candidates in steps 2-4 from 0.0700 to 0.0358 (48.9\%). Each prediction uses only earlier results. Candidate definitions and prediction diagnostics are given in the Supporting Information.

\FloatBarrier
\section{Conclusions}\label{sec:conclusions}

We have developed POMDP-QUBO for iterative "\emph{on-the-fly}" active-space selection at fixed electron and orbital counts. QUBO assembles candidate spaces from the available occupation, composition, and entanglement information. The POMDP selects the next calculation by considering its expected quality gain and the information it can provide. Calculated outcomes update the belief and guide subsequent choices, including when a valid candidate is rejected. The procedure refines orbital selection through calculations on the target molecule without requiring a pre-existing molecular training set.

The framework reproduces the expected valence active orbitals of small molecules, including bonding/anti-bonding and complementary pairs.
For the TiF$_3$, refinement within CAS(33e,25o) retains the five Ti~3d seed orbitals and brings both ligand-field excitation energies closer to the literature values. For 1,2-dioxetanone, the final CAS(16e,13o) spaces match the literature reference subspaces at the \smin{} and both conical intersections, \cioo\ and \cicc. These results demonstrate refinement of orbital content at a prescribed active-space size. 
Comparisons of the 2- and 3-channel schemes show that the 3-channel scheme recovers full reference coverage with fewer candidate optimizations (2 \textit{vs.} 4 at  \smin{}; 1 \textit{vs.} 6 at \cioo), 
while both routes reach 13/13 at  \smin{} with nearly identical energies; at \cioo\ the 2-channel CASSCF route reaches 12/13 within budget. 
Response learning and two-step planning further improve decisions: at a representative fourth-step \smin{} decision, the full model selects a candidate with lower immediate but higher two-step value, recovering 13/13 versus 12/13 for controls.

Automated active-space selection makes the POMDP-QUBO approach particularly well suited for integration with large language models, enabling the development of autonomous MC/MR computational agents.\cite{ferrag2026llm, chowa2026language} Such agents can help meet the urgent demand for high-throughput, high-quality theoretical data on complex strongly correlated systems,\cite{lischka2018multireference, dey2026high, zhang2026quantum, nguyen2026synthetic} including those in photochemistry, non-adiabatic processes, transition-metal catalysis, and lanthanide and actinide chemistry, especially, in the era of intelligent computing. We are actively pursuing this direction.

\begin{suppinfo}
Orbital scores and QUBO coefficients; Bayesian updates and candidate selection; computational settings; small-molecule valence checks; dioxetanone selection trajectories, state energies, and orbital overlaps; \smin{} decision controls and prediction errors; and TiF$_3$ state energies.
\end{suppinfo}

\section*{Author Contributions}
Chen Li implemented the algorithms, performed the quantum chemical calculations, analyzed the results, prepared the figures and tables, and wrote the first draft. Kaifan Pan contributed to algorithm testing and computational validation. Xinyu Sun and Yinxuan Song contributed to active-space analysis, orbital interpretation, and discussion of the chemical results. Lianhua He and Teng Lu contributed to method design, algorithm discussions, and result verification. Haibo Ma and Zhenggang Lan provided guidance on quantum chemistry and contributed to manuscript revision. Yingjin Ma conceived the study, guided method development, supervised the research, and revised the manuscript.

\section*{Conflict of Interest}
The authors declare no competing financial interest.

\section*{Acknowledgments}
This work was supported by Strategic Priority Research Program (XDB2100000),
National Natural Science Foundation of China (No.22673132, 22333003), and Project of Ganjiang Innovation Academy (E455F001) of Chinese Academy of Sciences.

\renewcommand{\bibfont}{\fontsize{9bp}{10bp}\selectfont}
\bibliography{references}
\end{document}